\documentclass[12pt]{article}
\usepackage{epsf}
\usepackage{amsmath}
\usepackage{graphics}

\renewcommand{\thefootnote}{\#\arabic{footnote}}

\newcommand{\lesssim}{ \mathop{}_{\textstyle \sim}^{\textstyle <} }

\begin{document}

\setcounter{footnote}{0}
\begin{titlepage}

\begin{center}

\hfill astro-ph/0505339\\
\hfill TU-744\\
\hfill May 2005\\

\vskip .5in

{\Large \bf
Implications of the Curvaton 
on Inflationary Cosmology
}

\vskip .45in

{\large
Takeo Moroi$^a$ and Tomo Takahashi$^b$
}

\vskip .45in

{\em
$^a$Department of Physics, Tohoku University\\
Sendai 980-8587, Japan \\
}

\vskip .2in

{\em
$^b$Institute for Cosmic Ray Research,
University of Tokyo\\
Kashiwa 277-8582, Japan
}

\end{center}

\vskip .4in

\begin{abstract} 
We study implications of the curvaton, a late-decaying light scalar
field, on inflationary cosmology, paying particular attentions to
modifications of observable quantities such as the scalar spectral
index of the primordial power spectrum and the tensor-to-scalar ratio.
We consider this issue from a general viewpoint and discuss how the
observable quantities are affected by the existence of the
curvaton. It is shown that the modification owing to the curvaton
depends on class of inflation models. We also study the effects of the
curvaton on inflation models generated by the inflationary flow
equation.
\end{abstract}

\end{titlepage}

\renewcommand{\thepage}{\arabic{page}}
\setcounter{page}{1}
\renewcommand{\thefootnote}{\#\arabic{footnote}}
\renewcommand{\theequation}{\thesection.\arabic{equation}}

\section{Introduction}
\setcounter{equation}{0}

Recent precise observations of the cosmic microwave background (CMB)
radiation \cite{Bennett:2003bz} and the large scale structure
\cite{Tegmark:2003uf} have provided deep insights into the origin of
the cosmic density fluctuations.  In particular, it has become clear
that the primordial density fluctuation is almost scale invariant.  As
a mechanism to generate the scale invariant density fluctuation,
inflation \cite{inflation} is the most prominent candidate.

Even with inflation, however, it is not automatic to make the present
density fluctuations to be consistent with the observational
constraints.  In the standard scenario, fluctuation of the inflaton
field, whose potential energy is responsible for the energy density
during the inflation, is generated during the inflation and it becomes
the origin of the present density fluctuations.  In this case,
properties of the present density fluctuations are determined once the
inflaton potential is fixed, and we obtain constraints on the
individual inflation models.

If there exists some scalar field other than the inflaton, however,
the mechanism of generating the cosmic density fluctuations may become 
more complicated.  In particular, with a scalar field whose mass is
much smaller than the expansion rate during the inflation, fluctuation 
is imprinted in the amplitude of such scalar field, which provides
another potential source of the present density fluctuations.  In this 
paper, we consider one of such examples, the curvaton
\cite{curvaton}.\footnote
{There is another mechanism producing the primordial fluctuation such
as inhomogeneous reheating or modulated reheating
\cite{Dvali:2003em,Kofman:2003nx}. In this paper, we do not consider
this possibility.}
In the curvaton scenario, there exists a late-decaying scalar
condensation (the curvaton) which acquires amplitude fluctuation
during the inflation.  When the curvaton decays, fluctuation of the
curvaton amplitude becomes the fluctuation of the radiation (and of
other components in the universe).  Thus, if the curvaton exists,
properties of the present density fluctuations change compared to the
standard scenario only with the inflaton.  From the particle physics
point of view, there are various well-motivated candidates of the
curvaton \cite{CurvCandidates}.

Since the recent results from Wilkinson microwave background probe
(WMAP) \cite{Bennett:2003bz} provided severe constraints on the
properties of the primordial density fluctuations, it is interesting
to reconsider the observational constraints in the framework of the
curvaton scenario.  With the curvaton, it is expected that the
constraints on the inflation models are changed (and possibly relaxed)
compared with the case without the curvaton.

Indeed, in the past, it was pointed out that the constraints on the
scale of the inflation is drastically relaxed with the curvaton
\cite{Dimopoulos:2002kt}.  Then, there have been subsequent works
focusing on the low scale inflation in the curvaton scenario
\cite{Lyth:2003dt,Matsuda:2003vj,Postma:2004an,Dimopoulos:2004yb}.
Furthermore, in Ref.\ \cite{Langlois:2004nn}, the authors has shown
that the quartic chaotic inflation model, which is marginally excluded
by WMAP observations, becomes viable when the curvaton mechanism
applies.  In Ref.\ \cite{Moroi:2005kz}, several other inflation models
were investigated such as chaotic inflation with several monomials,
the natural inflation model and the new inflation model.  The
discussions made in Refs.~\cite{Langlois:2004nn,Moroi:2005kz} are
mainly based on the scalar spectral index and the tensor-to-scalar
ratio.  Importantly, although some inflation models can become viable
with the curvaton even if they are excluded by current observations,
there exist other class of models which cannot be liberated even with
the curvaton. In particular, in Ref.~\cite{Moroi:2005kz}, it was
discussed how the liberation of inflation models depends on model
parameters such as the initial amplitude, mass and decay rate of the
curvaton field.

In this paper, we discuss the implication of the curvaton on inflation
models from some general points of view.  In the previous works, the
analysis have been rather dependent on inflaton potentials.  Of
course, it is important to study individual inflation models motivated
from, in particular, particle physics point of view.  It is, however,
also possible to adopt some other approach, parameterizing the
inflation models by using the slow-roll parameters (at some epoch
during the inflation).  In this approach, one generates inflation
models with some stochastic method such as using the inflationary flow
equations \cite{Hoffman:2000ue,Kinney:2002qn}.  Such a method has been
used in many studies including the one by the WMAP team and
constraints on inflation models have been discussed
\cite{Hansen:2001eu,Easther:2002rw,Peiris:2003ff,Kinney:2003uw,
Liddle:2003py,Chen:2004nx}.  Although other approaches such as
stochastic approaches or implementations of the flow equation are
proposed \cite{Ramirez:2005cy}, we discuss the implications of the
curvaton on inflation models generated by the usual flow equation as
an example.

The organization of this paper is as follows.  In the next section, we
start with briefly reviewing the formalism to study how observable
quantities are affected in the system containing the inflaton and the
curvaton.  We discuss how the standard scenario is modified in terms
of the background evolution and the perturbation.  We also review the
classification of single-field inflation models.  After the
preparation, in section 3, we go into how observational quantities are
affected by introducing the curvaton and discuss it from some general
viewpoints.  In section 4, we consider the effects of the curvaton on
inflation models generated by the inflationary flow equation.  Then we
conclude this note by summarizing the results in the final section.

\section{Formalism}

\subsection{Background evolution}

First we present the background evolution in the curvaton scenario
comparing its counterpart of the standard case.

In the standard case, the potential energy of the inflaton drives the
inflation. During inflation, the inflaton slowly rolls down the
potential. However, at some point, the inflaton starts to roll fast
down the potential hill, then the inflaton begins to oscillate around
the minimum of the potential.  When the potential of the inflaton can
be approximated by the quadratic form as $V(\chi) \sim
(1/2)m_\chi^2\chi^2$, the energy density of the inflaton behaves as
matter component. Then the universe is dominated by the oscillating
inflaton. We call this epoch ``$\chi$ dominated'' or ``$\chi$D.''
After some time, when the expansion rate of the universe, namely the
Hubble parameter, becomes as the same as the decay rate of the
inflaton, the inflaton decays into radiation. Then the universe
becomes radiation dominated. This is, what we call, the standard
thermal history of the universe.

When the curvaton exists, above picture is modified.  First we
consider the case where the initial amplitude of the curvaton is so
small that the potential energy of the curvaton does not drive the
(second) inflation.  During inflation, the curvaton field stays at
somewhere on the potential. If we adopt the potential of the curvaton
to be of the form $V(\phi) \sim (1/2)m_\phi^2 \phi^2$, the curvaton
begins to oscillate around the minimum of the potential when the
Hubble parameter becomes of the same order of the mass of the
curvaton.  During this epoch, the energy density of the curvaton
behaves as ordinary matter. This event usually happens when the
universe is dominated by radiation which comes from the decay of the
inflaton. Since the energy density of matter decreases slower than
that of radiation, at some point, the energy density of the curvaton
becomes to dominate the universe. We call this epoch as $\phi$
dominated or $\phi$D. After some time, the curvaton also decays into
radiation when the expansion rate of the universe becomes of the same order
as the decay rate of the curvaton. Then the universe becomes radiation
dominated again. We call this epoch as RD2 not to confuse with the
radiation dominated epoch from the inflaton decay. (We call the first
radiation dominated epoch as RD1.)

Next we discuss the case where the energy density of the curvaton can
drive the second inflation after the first one induced by the
inflaton. When the initial amplitude of the curvaton is large enough,
the second inflation can happen after the $\chi$D or RD1 epoch before
$\phi$ starts to oscillate. In this case, the universe experiences the
inflation era driven by the inflaton, $\chi$D era, RD1 era and the
second inflation driven by the curvaton potential energy, followed by
the $\phi$D era and RD2 era.  The second inflation can modify the
observable quantities drastically in some cases. We discuss this point
later.

\subsection{Density perturbations: Standard case}
Now we briefly review the issue of density perturbation in the
standard case.

To discuss observational consequences, we have to set up the initial
condition during radiation dominated era after the decay of the
inflaton.  For this purpose, we represent the primordial power
spectrum with the (Bardeen's) gravitational potential $\Phi$ which
appears in the perturbed metric in the conformal Newtonian (or
longitudinal) gauge as
\begin{equation}
  ds^2 = -a^2( 1 + 2\Phi) d\tau^2 + a^2 ( 1 - 2\Psi) dx^2,
\end{equation}
with $\tau$ being the conformal time.  The quantum fluctuation of the
inflaton $\delta\chi$ during inflation generates the curvature
perturbation as $\mathcal{R} = - (H/\dot{\chi}) \delta \chi$. Since
$\delta\chi\simeq H/2 \pi$ and, $\Phi$ and $\mathcal{R}$ are related
as $\Phi = -(2/3) \mathcal{R}$ during radiation-dominated era, the
power spectrum of the curvature perturbation from the inflaton
fluctuation is \cite{Stewart:1993bc}
\begin{equation}
    P_\Phi^{\rm (inf)} = 
    \left[ 1 - 2 (1+3 C) \epsilon_V + 2C \eta_V \right]
    \frac{1}{12 \pi^2 M_{\rm pl}^6} 
    \frac{V_{\rm inf}^3}{V_{\rm inf}^{'2}} \biggr|_{k=aH},
\end{equation}
where the ``prime'' represents the derivative with respect to $\chi$
and $C=-2+\ln 2+\gamma$, with $\gamma \simeq 0.577$ being the Euler's
constant.  Here, we used the ``potential'' slow-roll parameters.
There are two ways to define the slow-roll parameters: one is the one
using inflaton potential and the other is the one using the Hubble
parameter.  The ``potential'' slow-roll parameters are defined as
\begin{eqnarray}
  \epsilon_V \equiv \frac{1}{2} M_{\rm pl}^2 
  \left( \frac{V_{\rm inf}'}{V_{\rm inf}} \right)^2,
  ~~~
  \eta_V \equiv M_{\rm pl}^2 \frac{V_{\rm inf}''}{V_{\rm inf}},
  ~~~
  \xi^2_V \equiv M_{\rm pl}^2 
  \frac{V'_{\rm inf} {V^{'''}_{\rm inf} }}{V_{\rm inf}^2}.
\label{eq:slow_roll}
\end{eqnarray}
where the third parameter $\xi^2_V$ is considered to be the second order
in the slow-roll.

The ``Hubble'' slow-roll parameters are defined as
\begin{eqnarray}
\epsilon_H &\equiv &
 2 M_{\rm pl}^2 \left( \frac{H'}{H} \right)^2 = \epsilon_V, \notag \\
\eta_H &\equiv &  2 M_{\rm pl}^2 \frac{H^{''}}{H} 
= - \epsilon_V + \eta_V, \notag \\
\xi^2_H &\equiv & 2 M_{\rm pl}^4 \frac{H' H^{'''}}{H^2} 
= 
\xi_V^2 - 3 \epsilon_V \eta_V + 3 \epsilon_V^2
\label{eq:HubbleSR}
\end{eqnarray}
where we also write the relation between the Hubble and potential
slow-roll parameters to the first order in slow-roll.

The scalar spectral index is defined as
\begin{equation}
n_s -1 \equiv \frac{d \ln P_\Phi}{d \ln k}. 
\end{equation}
Using the slow-roll parameters, the spectral index can be written as
\begin{equation}
    n_s^{\rm (inf)} - 1 =  - 6 \epsilon_V + 2 \eta_V -
    2 (7+12C)\epsilon_V^2 + 2 (3+8C) \epsilon_V \eta_V - 2 C\xi_V^2,
    \label{eq:n_s}
\end{equation}
where ``(inf) '' represents that the expression of $n_s$ is for the
case where primordial fluctuation comes from the fluctuation of the
inflaton.

In the second order, we have the running of the scalar spectral index
which is
\begin{equation}
\frac{ d \ln n^{\rm (inf)}_s}{d \ln k} = -24 \epsilon_V^2 + 16 \epsilon_V \eta_V - 2 \xi_V^2.
\end{equation}

During inflation, the gravity wave can also be generated.  The
primordial gravity wave (tensor) power spectrum is given by
\begin{equation}
  P_T^{\rm (inf)}  = \frac{2V_{\rm inf}}{3 \pi^2 M_{\rm pl}^4}.
\end{equation}
With this expression, the tensor-to-scalar ratio $r$ is defined and given
by
\begin{equation}
    r^{\rm (inf)}
    \equiv 
    \frac{P_T^{\rm (inf)}}{P_\mathcal{R}^{\rm (inf)}}
    = 
    \frac{4}{9} \frac{P_T^{\rm (inf)}}{P_\Phi^{\rm (inf)}} 
    = 16 \epsilon_V \left( 1 + 4C \epsilon_V - 2 C\eta_V \right).
\end{equation}

\subsection{A classification of inflation models}

Here we discuss one of possible classifications of single-field
inflation models.  In many literatures, the following classification
has been used; single-field inflation models can be classified into
the ``small-field,'' `` large-field'' and ``hybrid-type'' models
\cite{Dodelson:1997hr}.  In this classification, the models are
classified on the $n_s - r$ plane and distinguished by the value of
the slow-roll parameters or the first and second derivatives of the
inflaton potential. To the first order in the slow-roll parameters,
the observational plane $(n_s, r)$ is uniquely divided into the three
classes.  In this section, we consider the slow-roll approximation in
the first order.

The ``small-field'' models are identified with the condition
$V''(\chi) < 0$ and $( V' /V)^2 > V''/V$, which also means $2
\epsilon_V > 0 > \eta_V$.  This category includes, for example, the
new and natural inflations. The generic potential of this type is
\begin{equation}
V(\chi) = \lambda v^4 \left[ 1 - \left( \frac{\chi}{v} \right)^p \right].
\end{equation}
The scalar spectral index and the tensor-to-scalar ratio of this type
of model can be written with the slow-roll parameter defined in
Eq.~(\ref{eq:slow_roll}) and the number of $e$-foldings during
inflation which is defined as $N_e \equiv \ln (a_{\rm end}/ a_* )$
where $a_{\rm end}$ and $a_*$ are the scale factor at the end of
inflation and the time of horizon crossing.  Using the slow-roll
approximation, $N_e$ is given by
\begin{equation}
N_e = \frac{1}{M_{\rm pl}^2} \int_{\chi_{\rm end}}^{\chi_*} 
\frac{V_{\rm inf}}{V'_{\rm inf}} d \chi.
\end{equation}
Using the $e$-folding number $N_e$, the slow-roll parameters for this
type of model are given by, for $p>2$,
\begin{eqnarray}
\epsilon_V 
&=&
\frac{p^2}{2} \left( \frac{M_{\rm pl}}{v} \right)^2 
\left[ \frac{1}{p(p-1) N_e} \left( \frac{v}{M_{\rm pl}} \right)^2 
\right]^{\frac{2 (p-1)}{p-2}},
\\
\eta_V 
&=&
- \frac{p-1}{p-2} \frac{1}{N_e}
\end{eqnarray}
For $v < M_{\rm pl}$, the $\epsilon$ parameter becomes very small,
thus the spectral index is given by
\begin{equation}
n_s^{\rm (inf)} -1 = - 2 \frac{p-1}{p-2} \frac{1}{N_e}
\label{eq:small_ns}
\end{equation}

The ``large-field models'' are identified with the condition
$V''(\chi) > 0$ and $( V' /V)^2 > V''/V$, which also means 
$2 \epsilon_V> \eta_V > 0$. This category includes, for example,
the chaotic inflation.  The generic potential of this type is
\begin{equation}
    V(\chi) = \lambda M_{\rm pl}^4 
    \left( \frac{\chi}{M_{\rm pl}} \right)^\alpha.
    \label{eq:V_large}
\end{equation}
The $n_s$ and $r$ are obtained as similarly as the case with the
small-field models as
\begin{eqnarray}
n_s^{\rm (inf)}-1 &=& -\frac{\alpha + 2 }{2 N_e}, 
\label{eq:large_ns} \\
r^{\rm (inf)} &=& \frac{4\alpha}{N_e}.
\end{eqnarray}
Thus inflation models in this category predict red-tilted scalar
primordial spectrum and relatively large tensor-to-scalar ratio.

The final one, ``hybrid-type'' models are identified with the
condition $V''(\chi) > 0 $ and $( V' /V)^2 < V''/V$, which implies
$2 \epsilon_V < \eta_V$. This category includes, the
hybrid inflation model as the name tells. Only models of this category
can give blue-tilted spectrum.

\subsection{Effects of the curvaton fluctuation}

Here we briefly review the effects of the fluctuation of the curvaton
on the observable quantities such as $n_s$ and $r$.  (For details, see
\cite{Langlois:2004nn,Moroi:2005kz,Moroi:2002rd}.)  For the case where the
fluctuations of the inflaton and the curvaton both affect the cosmic
density perturbation today, the gravitational potential during RD2 era
is given by
\begin{equation}
  \Phi_{\rm RD2}
  = 
  - \left[ 1 - (1+3C) \epsilon_V + C\eta_V \right]
  \frac{2}{3M_{\rm pl}^2} \frac{V_{\rm inf}}{V_{\rm inf}'}  
  \delta \chi_{\rm init}
  - f(X) \frac{\delta \phi_{\rm init}}{M_{\rm pl} },
  \label{eq:Phi_mod}
\end{equation}
where $\delta\phi_{\rm init}$ is the primordial fluctuation of the
curvaton, and
\begin{eqnarray}
  X = \frac{\phi_{\rm init}}{M_{\rm pl}}.
\end{eqnarray}
The function $f(X)$ represents the size of the contribution from the
fluctuation of the curvaton. $f(X)$ can be calculated using the linear
perturbation theory \cite{Langlois:2004nn}.  Assuming that the
curvaton dominates the universe once in the course of the history of
the universe, $f(X)$ can be given, for small and large $X$, by
\begin{eqnarray}
    f(X) \simeq \left\{
        \begin{array}{ll}
            \displaystyle{ \frac{4}{9X}} 
            & ~~~:~~~ \phi_{\rm init} \ll M_{\rm pl}
            \\ \\
            \displaystyle{ \frac{1}{3}X}
            & ~~~:~~~ \phi_{\rm init} \gg M_{\rm pl}
        \end{array} \right. .
    \label{funf}
\end{eqnarray}

Using Eq.\ (\ref{eq:Phi_mod}), the scalar spectral index can be written
as
\begin{equation}
  P_\Phi = 
  \left[
  1 + \tilde{f}^2 (X) \epsilon_V 
  -2(1+3C)\epsilon_V + 2 C\eta_V 
  \right]
  \frac{V_{\rm inf}}{54 \pi^2 M_{\rm pl}^4 \epsilon_V},
\end{equation}
where $\tilde{f} = (3/\sqrt{2}) f$.  Thus the scalar spectral index is
given by using Eq.~(\ref{eq:n_s})\footnote
{Here we assume that the mass of the curvaton is much smaller than
that of the inflaton. Thus we can neglect the term in the expression
of $n_s$ which comes from the curvature of the curvaton potential as
$\pm 2 \eta_{\phi \phi}$ where the positive and negative signs are for
the case with $X<1$ and $X>1$ respectively and
\begin{equation*}
\eta_{\phi \phi} = M_{\rm pl}^2 \frac{m_{\phi}^2}{V_{\rm total}}.
\end{equation*}
Here $V_{\rm total} = V_{\rm inf} + V_\phi$.}
\begin{equation}
  n_s-1 =  
  - 2 \epsilon_V 
  +  \frac{ 2 \eta_V - 4 \epsilon_V }{1 + \tilde{f}^2 \epsilon_V} 
  - 2 (7+12 C )\epsilon_V^2 + 2 (3+8C) \epsilon_V \eta_V - 2 C\xi^2_V 
  \label{eq:ns_mod}
\end{equation}
The running of the scalar spectral index is given by 
\begin{equation}
\frac{d \ln n_s}{d \ln k} = 4 \epsilon_V (\eta_V - 2 \epsilon_V ) 
- \frac{ 16 \epsilon^2_V -12 \epsilon_V \eta_V + 2 \xi_V^2 }{ 1 + \tilde{f}^2 \epsilon_V }
+ \frac{4 \tilde{f}^2 \epsilon_V (\eta_V - 2 \epsilon_V)^2 }{(1 + \tilde{f}^2 \epsilon_V)^2 }.
\label{eq:dndk_mod}
\end{equation}

The tensor power spectrum is not modified even with the
curvaton. However, since the scalar perturbation spectrum is modified,
the tensor-to-scalar ratio becomes
\begin{equation}
  r = \frac{16 \epsilon} {1 + \tilde{f}^2 \epsilon}
        ( 1 + 4C\epsilon_V - 2 C\eta_V ).
  \label{eq:r_mod}
\end{equation}

\subsection{Effects of modification of the background evolution}
\label{sec:background}

Here we discuss effects of the modification of the background
evolution due to the curvaton.  Generally, the observable quantities
such as the scalar spectral index and the tensor-to-scalar ratio
depends on the field value of the inflaton at the time of horizon
crossing which can usually be expressed with the number of
$e$-foldings during inflation.  The fluctuation which corresponds to
some reference scale $k_{\rm ref}$ at present epoch can be
approximately given, in the standard case, as \cite{Liddle_Lyth}
\begin{equation}
\frac{k_{\rm ref}}{a_0 H_0} 
=
\frac{a_k H_k}{a_0 H_0}
= 
e^{-N_e(k)}
 \frac{a_{\rm end}}{a_{\rm reh}} 
 \frac{a_{\rm reh}}{a_0}
\frac{H_k}{H_0}
\label{eq:standard}
\end{equation}
where ``end'' and ``reh'' denote the time when the inflation ends and
the reheating epoch, i.e., the beginning of RD epoch.  Thus the
$e$-folding number during inflation is given by
\begin{equation}
N_e(k)^{\rm (standard)} \simeq
- \ln \frac{k}{a_0 H_0} 
- \frac{n+2}{6n} \ln \frac{\rho_{\rm end}}{\rho_{\rm reh}}
+ \frac{1}{3} \ln \frac{s_0}{s_{\rm reh}}
+ \ln  \frac{H_{\rm inf}}{H_0}
\end{equation}
where we assumed that the energy density of the oscillating inflaton
behaves as $\rho_\chi \propto a^{-6n/(n+2)}$ which corresponds to the
case with inflaton potential $V_{\rm inf} \propto \chi^n$.  $s_{\rm
reh}$ and $s_0$ are the entropy density at present time and time of
reheating, respectively.  In the last term, we assumed that the Hubble
parameter is almost constant during inflation, thus we replaced $H_k$
with $H_{\rm inf}$.  Notice that the decay rate of the inflaton
$\Gamma_\chi$ and parameters in the potential are needed to determine
$N_e$ exactly. Depending on $\Gamma_\chi$ and the parameters in the
potential, the number of $e$-folding can be changed as
\begin{equation}
N_e \simeq \frac{4 -\alpha}{6\alpha} \ln \Gamma_\chi + \cdots
\end{equation}

Now we discuss the $e$-folding number in the curvaton scenario.  When
the curvaton is introduced, Eq.\ (\ref{eq:standard}) is modified as
\begin{equation}
\frac{k_{\rm ref}}{a_0 H_0} 
= 
e^{-N_e(k)^{\rm (curvaton)}}
\frac{a_{\rm end}}{a_{\rm reh1}} 
\frac{a_{\rm reh1}}{a_{\rm inf2}}
\frac{a_{\rm inf2}}{a_{ \phi \rm D}}
\frac{a_{\phi \rm D}}{a_{\rm reh2}}
\frac{a_{\rm reh2}}{a_0}
\frac{H_k}{H_0}
\end{equation}
where ``reh1,'' ``inf2,'' ``$\phi$D2,'' and ``reh2'' denote the time
when the first radiation dominated epoch begins, the second inflation
begins, the oscillating curvaton dominated begins and the second
radiation dominated epoch begins, respectively.  When the initial
amplitude of the curvaton is small ($\phi_{\rm init} \ll M_{\rm pl}$),
there is no second inflation. In that case, $(a_{\rm reh1}/a_{\rm
inf2}) (a_{\rm inf2}/a_{ \phi \rm D})$ should be replaced with
$(a_{\rm reh1}/a_{\phi \rm D})$.

\begin{figure}[t]
    \begin{center}
        \centerline{\epsfysize=0.75\textwidth\epsfbox{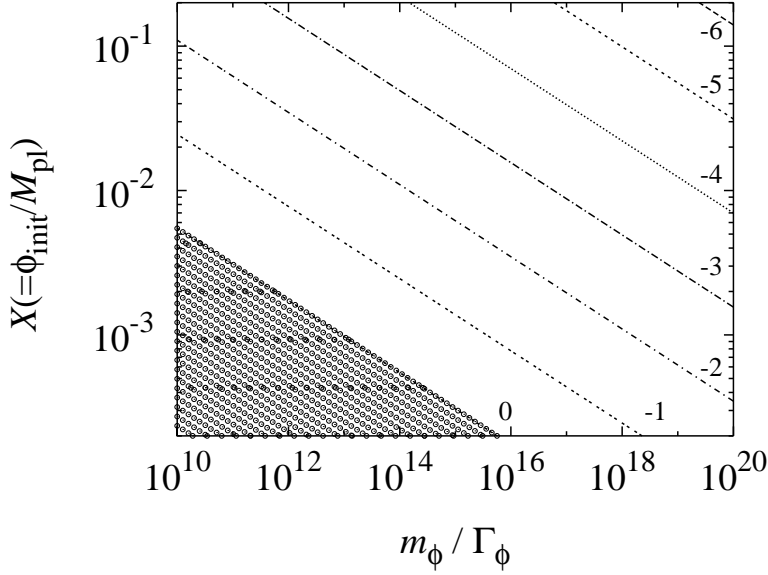}}
        \caption{Contours of constant 
        $\Delta N_e$ in the $m_\phi / \Gamma_\phi$ vs.\ $X$ plane.  The
        region where the curvaton does not dominate the universe is
        represented with small circles.}
        \label{fig:delta_N}
\end{center}
\end{figure}

Assuming that the curvaton field 
does not dominate the energy density of the universe when 
the curvaton begins to oscillate,\footnote
{This assumption almost corresponds to the condition $\phi_{\rm init}
\le M_{\rm pl}$.}
namely the case with no second inflation driven by the curvaton,
we get \cite{Liddle:2003as}.
\begin{eqnarray}
    N_e(k)^{\rm (curvaton)} 
    &\simeq& 
    N_e(k)^{\rm (standard)}
    - \frac{1}{12} \ln \frac{\rho_{\phi \rm D}}{\rho_{\rm reh2}}
    \notag \\
    &\simeq&
    N_e(k)^{\rm (standard)}
    - \frac{1}{6} \ln \frac{m_\phi}{\Gamma_\phi}
    - \frac{2}{3} \ln \frac{\phi_{\rm init}}{M_{\rm pl}}
\end{eqnarray}
In Fig.~\ref{fig:delta_N}, we plot contours of constant $\Delta N_e
\equiv N_e^{\rm (curvaton)} - N_e^{\rm (standard)}$ in the $m_\phi /
\Gamma_\phi$ vs.\ $X$ plane.  Requiring that the curvaton dominates
the energy density of the universe after the first radiation dominated
epoch, i.e., $\rho_\phi > \rho_{\rm rad}$ at the time when the
curvaton decays into radiation, we have the relation among the mass,
the decay rate and the initial amplitude of the curvaton as
\begin{equation}
\frac{m_\phi}{\Gamma_\phi} > \frac{1}{9} X^4.
\end{equation}
In Fig.~\ref{fig:delta_N}, the region where the above inequality is not
satisfied is represented with small circles.

As seen from Eqs.~(\ref{eq:small_ns}) and (\ref{eq:large_ns}), the
scalar spectral index in the small-field and large-field models are
inversely proportional to $N_e$.  Thus slight change of $N_e$ affects
$n_s^{\rm (inf)}$ as $|\Delta n_s^{\rm (inf)}| \sim \Delta N_e /N_e^2$.
This implies that, requiring $|\Delta n_s^{\rm (inf)}| < 0.001$,
$\Delta N_e < 4 - 5$ for $N_e = 50 - 60.$ Notice that the uncertainty
of the $e$-folding number $N_e$ also comes from the inflaton sector
through the decay rate of the inflaton and the form of the potential,
in particular around the minimum (i.e., the potential which the
inflaton feels when the inflaton oscillates.)

When the second inflation occurs, the number of $e$-foldings can be
written as
\begin{eqnarray}
    N_e(k)^{\rm (curvaton)} 
    &\simeq& 
    N_e(k)^{\rm (standard)}
    - N_2
    - \frac{1}{12} \ln \frac{\rho_{\phi \rm D}}{\rho_{\rm reh2}}
    \notag \\
    &\simeq& 
    N_e(k)^{\rm (standard)}
    - N_2 
    - \frac{1}{6} \ln \frac{m_\phi}{\Gamma_\phi}
    - \frac{1}{6} \ln \frac{\phi_{\rm init}}{M_{\rm pl}}
\end{eqnarray}
where $N_2$ represents the $e$-folding number during the second
inflation which can be $20-30$ \cite{Moroi:2005kz}.  Thus, in this
case, $N_e$ drastically reduced due to the second inflation driven by
the curvaton, which significantly affects the observational
quantities.

\section{Effects on the Observable Quantities}

Now we discuss how the curvaton affects the quantities such as $n_s$
and $r$. In this section, we assume that the initial amplitude of the
curvaton is small compared with the Planck mass and the change of the
number of $e$-foldings is also small.  Furthermore, we denote the
scalar spectral index for the standard case (i.e., the case with the
inflaton fluctuation only) as $n_s^{\rm (inf)}$ and that for the case
with the inflaton and the curvaton as $n_s$.  In this section, we
consider the slow-roll in the first order.

\begin{figure}[t]
    \begin{center}
        \centerline{\epsfysize=0.75\textwidth\epsfbox{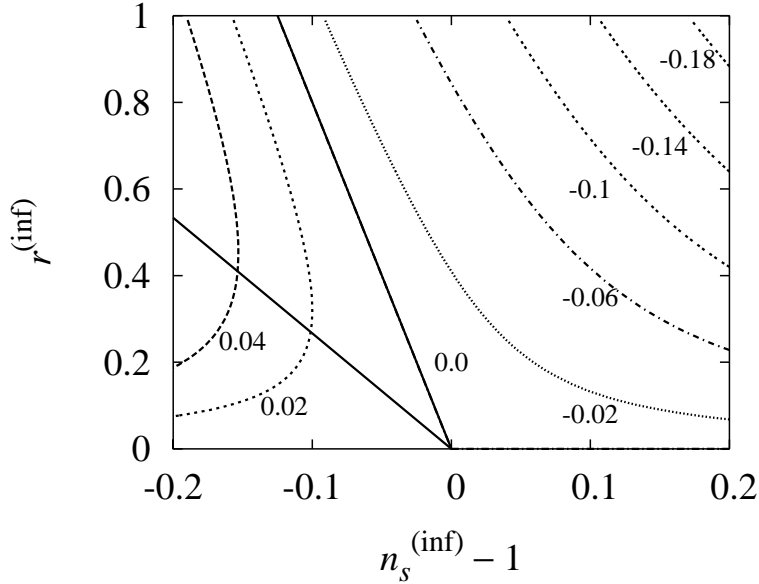}}
        \caption{Contours of constant $n_s - n_s^{\rm (inf)}$
        in the $n_s^{\rm (inf.)} -1$ vs.\ $r^{\rm (inf)}$ plane.  For
        reference, boundaries of ``small-field,'' ``large-field'' and
        ``hybrid-type'' models are also shown.  Notice that the
        boundary between the ``large-field'' model and ``hybrid-type''
        model coincides with the contour of $n_s - n_s^{\rm (inf)}=0$.
        We fixed the size of the contribution from the curvaton
        fluctuation as $\tilde{f}=5$.}
        \label{fig:ns}
    \end{center}
\end{figure}

First, we discuss to what extent the scalar spectral index is modified
when the curvaton comes into play. For this purpose, we show contours
of constant $n_s - n_s^{\rm (inf)}$ in the $n_s^{\rm (inf)} -1$ vs.\
$r^{\rm (inf)}$ plane in Fig.\ \ref{fig:ns}.  In the figure, we fix
contribution from the fluctuation of the curvaton as $\tilde{f}=5$
which corresponds to the case with $\phi_{\rm init} \sim 0.1 M_{\rm
pl}$. Interestingly, in the ``hybrid-type'' models, the scalar
spectral index always decreases, on the other hand, in the
``small-filed'' and ``large-field'' models, $n_s$ increases. This is
easily understood from Eq.\ (\ref{eq:ns_mod}).  When $2 \eta_V - 4
\epsilon_V$ is positive, which is the boundary of the ``hybrid-type''
and ``large-field'' models, the contribution from the curvaton always
suppresses the second term in Eq.\ (\ref{eq:ns_mod}). Thus $n_s$
becomes smaller. When $2 \eta_V - 4 \epsilon_V$ is negative, which is
the case for ``small-field'' and ``large-field'' models, the opposite
happens, namely $n_s$ becomes larger.  In addition, as seen from Eq.\
(\ref{eq:ns_mod}), the contribution from the curvaton fluctuation
comes in the combination of $\epsilon_V \tilde{f}^2$. Thus models with
small $\epsilon_V$ cannot be affected by the curvaton fluctuation
much.  The ``small-field'' model mostly covers a region where the
$\epsilon_V$ parameter is small.  Thus, as for the scalar spectral
index, we can observe the following feature for each class of
inflation models: In the hybrid-type inflation models, the spectral
index always decreases. In the large-field inflation models, the
scalar spectral index always increases. In the small-field inflation
models, the scalar spectral index increases very slightly, and this
class of models are not significantly affected by the existence of the
curvaton much.

\begin{figure}
    \begin{center}
        \centerline{\epsfysize=0.5\textwidth\epsfbox{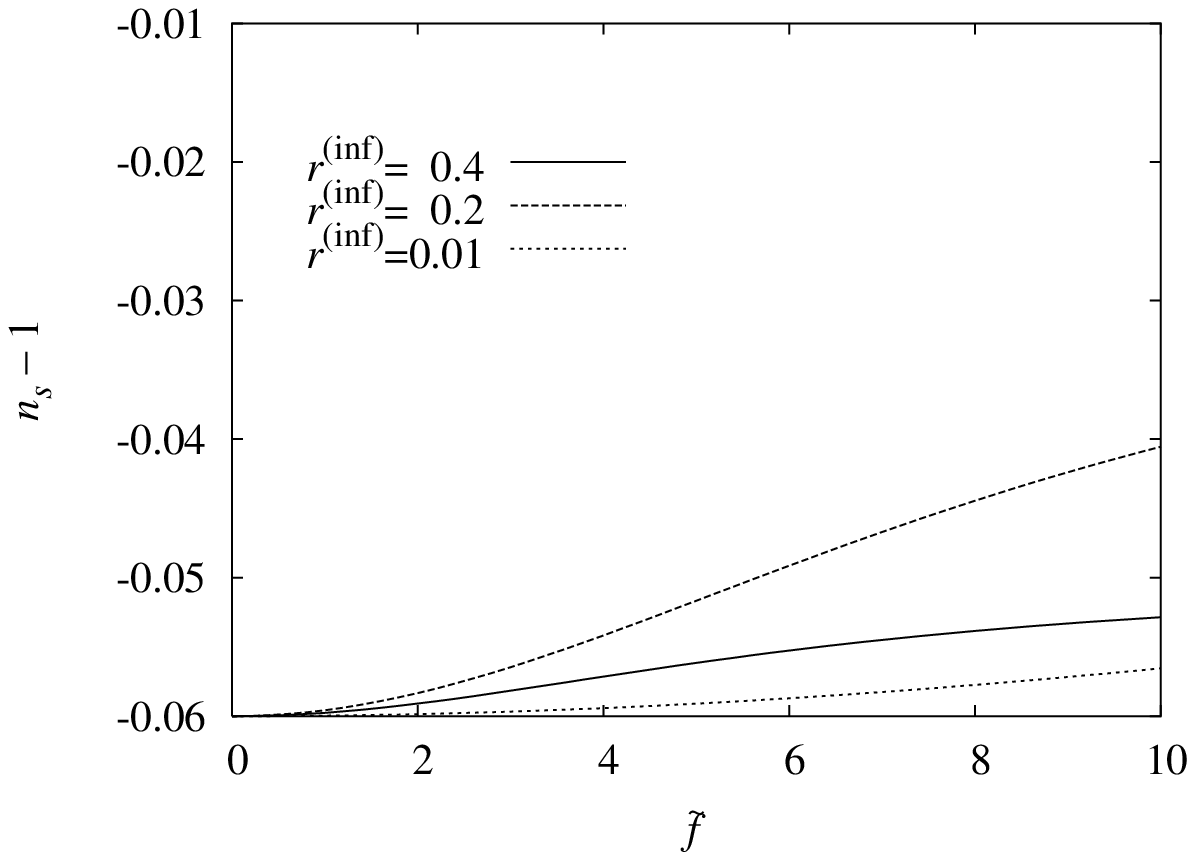}}
        \caption{$n_s-1$ as a 
        function of $\tilde{f}$ for $n_s^{\rm (inf)}-1=-0.06$.  Here,
        we take $r^{\rm (inf)}=0.4$ (solid line), $0.2$ (dashed line)
        and $0.01$ (dotted line).}
        \label{fig:n-0.06}
    \end{center}
    \begin{center}
        \centerline{\epsfysize=0.5\textwidth\epsfbox{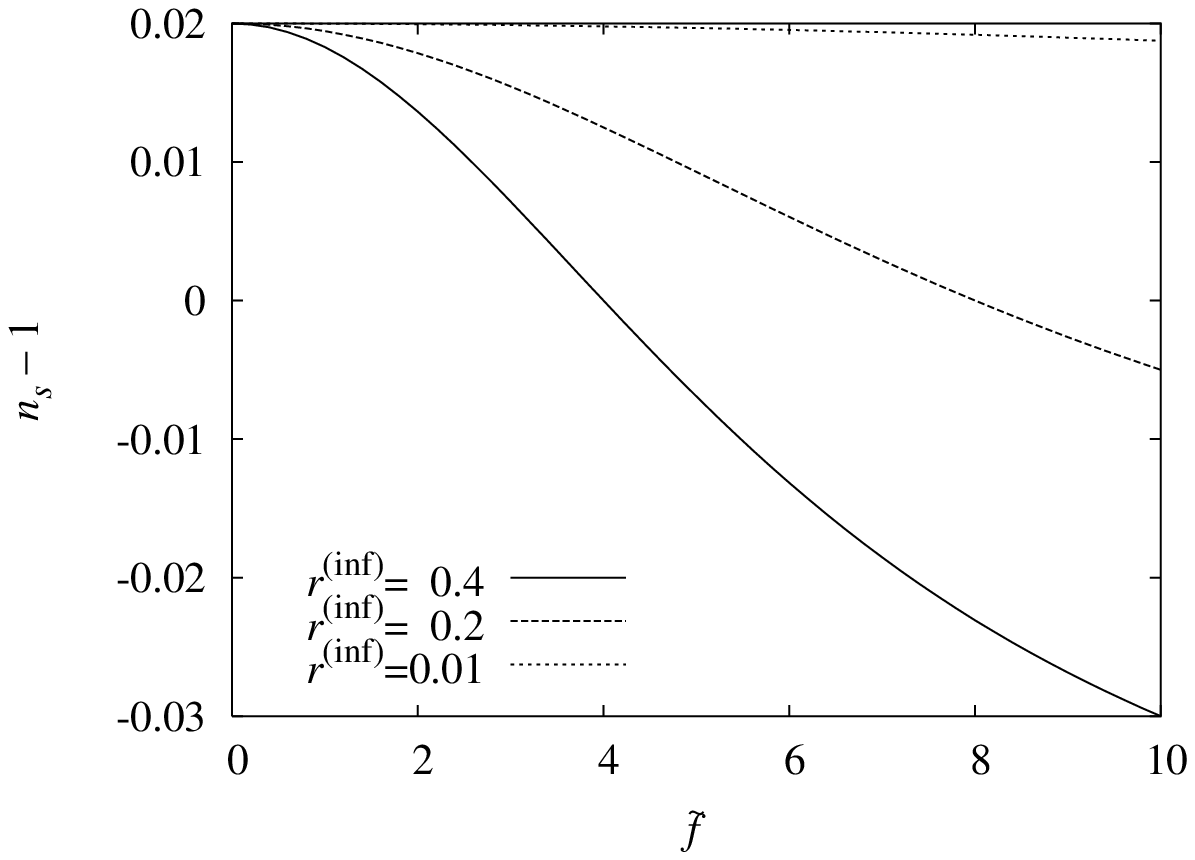}}
        \caption{Same as Fig.\ \ref{fig:n-0.06}, except for 
        $n_s^{\rm (inf)}-1=0.02$.}
        \label{fig:n+0.02}
    \end{center}
\end{figure}

Next we discuss $\tilde{f}$ dependence of the scalar spectral index.
In Figs. \ref{fig:n-0.06} and \ref{fig:n+0.02}, the changes of 
$n_s-1$ are shown as a function of $\tilde{f}$.
In Fig.~\ref{fig:n-0.06}, the case with the ``small-field'' model
($r^{\rm (inf)}=0.01$) and the ``large-field'' case ($r^{\rm
(inf)}=0.2$ and $0.4$) are shown.  As seen from the figure, for
``small-field'' model, (i.e., the case with small $r^{\rm (inf)}$),
the change of $n_s$ is small compared to the case with the
``large-field'' models.  In Fig.~\ref{fig:n+0.02}, the case with the
``hybrid-type'' are shown.  We can see from these figure that cases
with small $r^{\rm (inf)}$ (or $\epsilon$) shows slow response to the
increase in $f$  as mentioned above.

\begin{figure}[t]
    \begin{center}
        \centerline{\epsfysize=0.5\textwidth\epsfbox{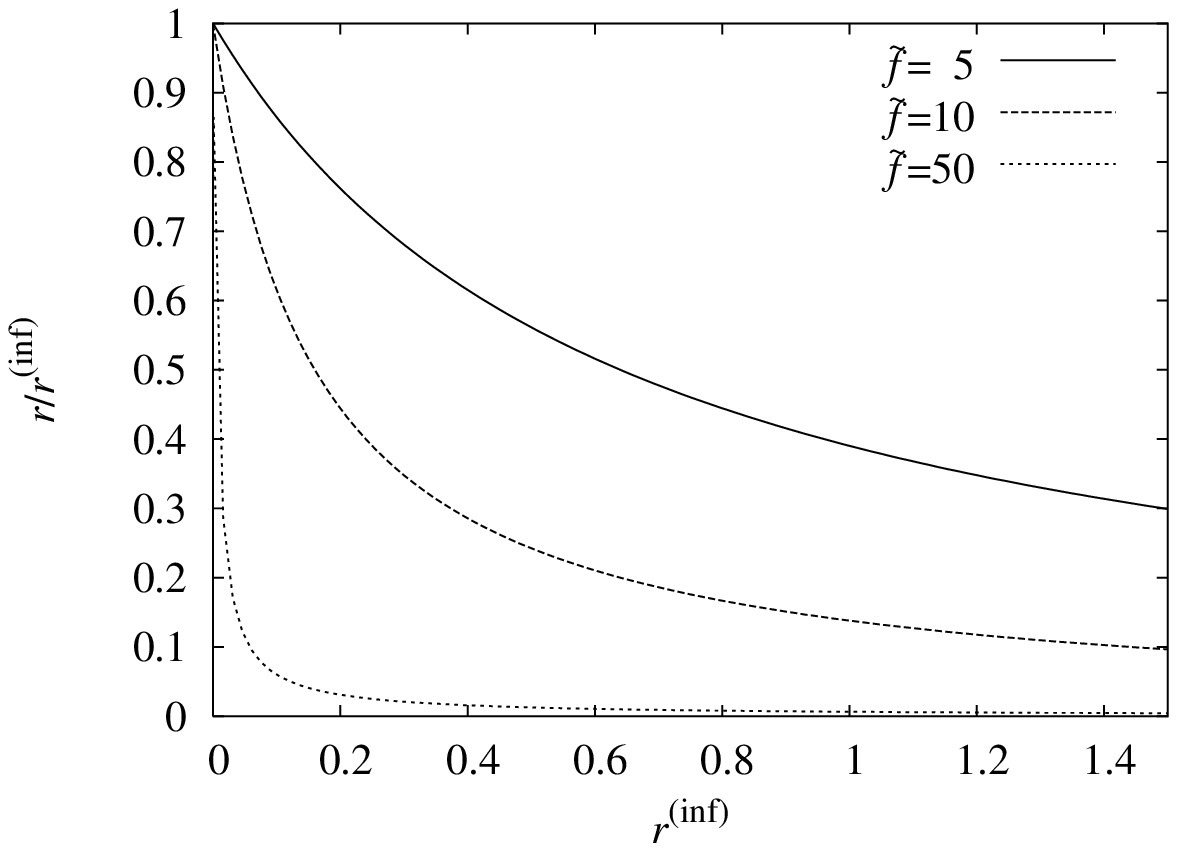}}
        \caption{The ratio $r/r^{\rm (inf)}$ as a function of 
        $r^{\rm (inf)}$.  Here, we take
        $\tilde{f}=5$, $10$, and $50$.}
        \label{fig:r}
    \end{center}
\end{figure}

Next we consider how the tensor-to-scalar ratio is affected by the
existence of the curvaton. For this purpose, in Fig.\ \ref{fig:r}, we
plot $ r/r^{\rm (inf)}$ as a function of $r^{\rm (inf)}$ for several
values of $\tilde{f}$.  As easily seen from Eq.\ (\ref{eq:r_mod}), the
effects of the curvaton always suppress the tensor-to-scalar ratio.
Furthermore, its effect is larger when the $\epsilon_V$ parameter
(i.e., $r^{\rm (inf)}$) is large since the contribution from the
curvaton comes in the form of $\epsilon_V \tilde{f}^2$ as discussed
above.

Here we comment on the consistency relation of the inflationary
quantities.  In the standard single inflaton case, we have the
consistency relation which is
\begin{equation}
r^{\rm (inf)} =  - 8 n_T.
\end{equation}
In the curvaton scenario, this equation is modified as
\cite{Langlois:2004nn}
\begin{equation}
r = \frac{- 8 n_T}{ 1 - {\tilde{f}}^2  n_T  / 2}.
\label{eq:cons_curvaton}
\end{equation}
Thus the consistency relation is also modified.  This can be used to
differentiate between the standard and curvaton scenarios.  It is also
important to notice that Eq.~(\ref{eq:cons_curvaton}) can be used to
obtain $\tilde{f}$, in other words, the initial amplitude of the
curvaton since $\tilde{f}$ is determined once $n_T$ and $r$ are
observationally determined.

\section{Analysis with the Flow Equation}

In this section, we discuss the effects of the curvaton on inflation
models using the flow equation approach \cite{Hoffman:2000ue}, which
is widely accepted to study some aspects of generic inflation models
\cite{Kinney:2002qn,Hansen:2001eu,Easther:2002rw,
Peiris:2003ff,Kinney:2003uw,Liddle:2003py,Chen:2004nx,Ramirez:2005cy}.
Before we consider the effects of the curvaton, first we briefly
review this approach, following Ref.~\cite{Kinney:2002qn}.

\subsection{Flow equation}

In this section, we make use of the Hubble slow-roll parameters which
are defined in Eqs.~(\ref{eq:HubbleSR}).  We can extend the Hubble
slow-roll parameters up to an arbitrary order as
\begin{equation}
{^l\lambda_H} \equiv
 \left( 2 M_{\rm pl}^2 \right)^l
\frac{1}{H^l} \left( \frac{dH}{d\chi} \right)^{l-1} 
\frac{d^{(l+1)} H}{d\chi^{(l+1)}}~;~~~ l \ge 1.
\end{equation}
In fact, ${^1\lambda_H}$ and ${^2\lambda_H}$ correspond to $\eta_H$
and $\xi_H$ respectively.  Using the definitions of the Hubble
slow-roll parameters and
\begin{equation}
\frac{d}{dN} = \sqrt{2} M_{\rm pl} \sqrt{\epsilon_H} \frac{d}{d\chi},
\end{equation}
we can obtain differential equations which are called ``inflationary
flow'' equation as
\begin{eqnarray}
\frac{d \epsilon_H}{dN} 
&=&
2 \epsilon_H (\eta_H - \epsilon_H ),  \notag \\
\frac{d\sigma_H}{dN} 
&=& 
-5 \epsilon_H \sigma_H - 12 \epsilon_H^2
 + \left( {^2\lambda_H} \right)^2, \notag \\
\frac{d ( {^l\lambda_H} )}{dN} 
&=& 
\left[ \frac{l-1}{2} \sigma_H + (l-2) \epsilon_H \right] {^l\lambda_H}
+ {^{l+1}\lambda_H} ~;~~ l \ge 2.
\label{eq:flow}
\end{eqnarray}
Here we defined $\sigma_H = 2 \eta_H - 4 \epsilon_H$ for convenience.
It is known that there are two classes of fixed points in the system
of the equations (\ref{eq:flow}) \cite{Kinney:2002qn}. The first class
is the case with $\epsilon_H= {^l\lambda_H}=0$ and $\sigma_H =$
constant. The second one is the case with $\sigma_H = -2
\epsilon_H=$constant and $ {^2\lambda_H} = \epsilon_H^2$, which
corresponds to the case with the power-law inflation.

Using the flow equation, we computed 50000 realizations of inflation
models following the method of Ref.~\cite{Kinney:2002qn}.  We
truncated the hierarchy at the fifth order to have a finite set of
equations for numerical calculations by setting $ {^l\lambda_H}=0$ for
$l\ge 6$.  To generate inflation models, we randomly choose the
initial conditions for the slow-roll parameters as
\begin{eqnarray}
    0 \leq \epsilon_{H} \leq 0.8,~~~
    -0.5 \leq \sigma_{H} \leq 0.5,
\end{eqnarray}
and
\begin{eqnarray}
    - 5\times 10^{-l} \leq {^l\lambda_{H}} \leq
    5\times 10^{-l} ~~~ (l=2, 3, 4, 5),
\end{eqnarray}
reducing the width of the range of the parameters by factor of ten for
each higher order in slow roll.  For $l\ge 6$, we approximate
${^l\lambda_{H}}=0$.

We also have to set the range of the number of $e$-folding during
inflation to generate inflation models.  In most analysis done so far
\cite{Kinney:2002qn,Easther:2002rw,Peiris:2003ff,Kinney:2003uw}, the
following range is used $40\leq N_e\leq 70$.  However, since we are
going to consider the effects of the curvaton on the models generated
by this method and we have discussed that the $e$-folding number can
be changed due to the existence of the curvaton in the previous
section, we need to set this quantity taking the modification by the
curvaton into account.

\subsection{Effects of the curvaton}

Now we discuss the effects of the curvaton on inflation models
generated by this method.  Since the fluctuation of the inflaton and
the curvaton are independent, we can calculate the observational
quantities according to Eqs.~(\ref{eq:ns_mod}) and (\ref{eq:r_mod}).
As mentioned above, we have to set the range of the $e$-folding number
taking into account the modification of the background evolution due
to the curvaton.  First we consider the case with small curvaton
initial amplitude (i,e., $X \ll 1$).  In this case, as discussed in
the section \ref{sec:background}, the modification of the background
evolution is relatively small.  Thus, in generating the inflation
models, we adopt the original range of the $e$-folding number $40\leq
N_e\leq 70$ even for the case with the curvaton.  

\begin{figure}
    \begin{center}
        \centerline{\epsfysize=0.45\textwidth
        \epsfbox{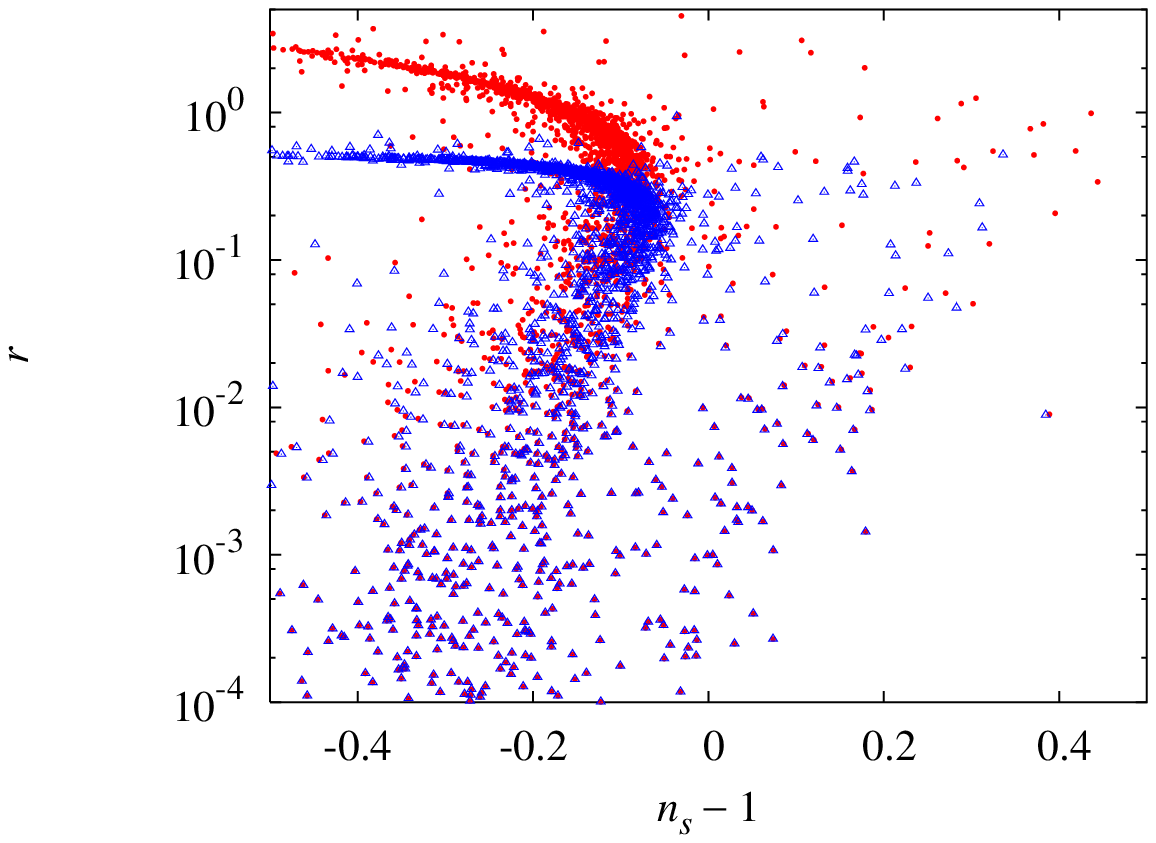}}
        \caption{Inflation models generated by the flow equation in
        the $n_s -1$ vs.\ $r$ plane.  Models in the standard scenario
        are shown in red (circles) while the curvaton cases with
        $\tilde{f}=5$ are shown in blue (triangles).}
        \label{fig:flow_ns_r}
    \end{center}
    \begin{center}
        \centerline{\epsfysize=0.45\textwidth
        \epsfbox{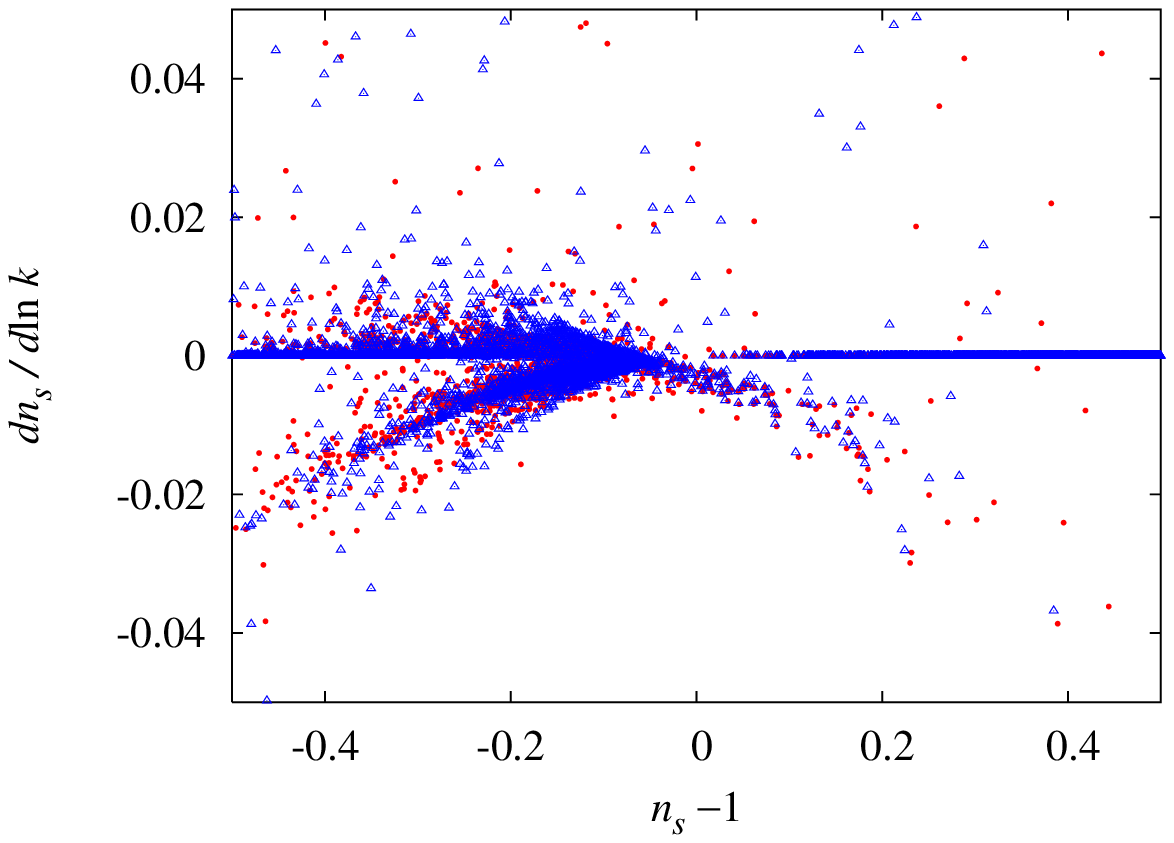}}
        \caption{Inflation models generated by the flow equation 
        in the $n_s -1$ vs.\ $d \ln n_s/dk$ plane.  Models in the
        standard and curvaton scenarios are shown in red (circles) and
        blue (triangles) respectively. }
        \label{fig:flow_ns_run}
    \end{center}
\end{figure}

In Fig.~\ref{fig:flow_ns_r}, we show how the distribution on the $n_s$
vs.\ $r$ plane is modified.  In the figure, distribution for the
standard case (without the curvaton) is shown in red color (or
circles) while that in the curvaton scenario with $\tilde{f}=5$ is
indicated in blue color (or triangles).  We can see some clustering
structures of generated models near the the second class of the fixed
point for the standard case.  Importantly, most of such clustering
region is excluded by the WMAP data because of too small $n_s$ and/or
too large $r$.  With the curvaton, the clustering occurs at the region
with smaller value of $r$.  Since $r$ always decreases in the curvaton
scenario, the cluster is shifted to lower $r$ region with curvaton.
However, the change of the spectral index due to the curvaton is not
significant enough and hence, if we take $40\leq N_e\leq 70$, the
large class of generated inflation models are excluded falling into
the clustering region.

In Fig.~\ref{fig:flow_ns_run}, we also show the distribution on the
$n_s$ vs.\ $d\ln n_s/dk$ plane.  As one can see, running of the index
$d\ln n_s/dk$ is quite small even with the curvaton.

\begin{figure}
    \begin{center}
        \centerline{\epsfysize=0.45\textwidth
        \epsfbox{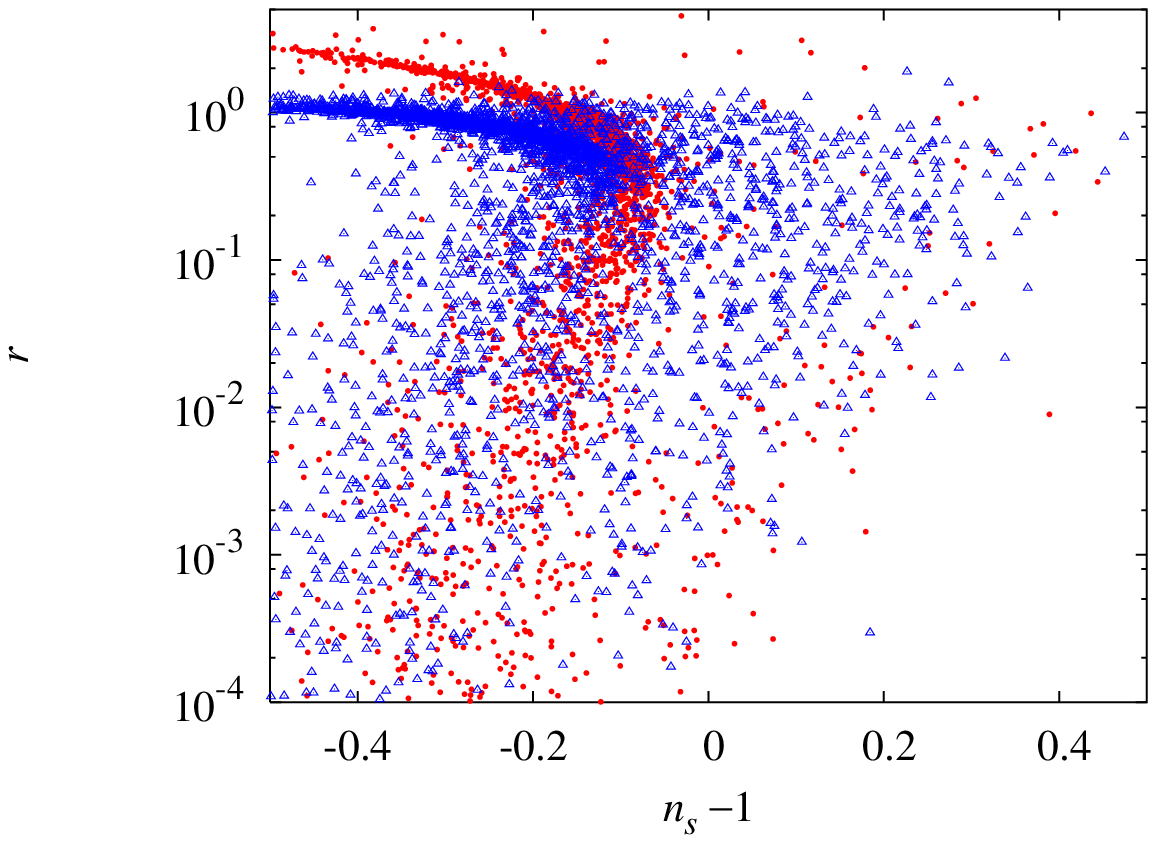}}
        \caption{The same as Fig.~\ref{fig:flow_ns_r} except
        that we take $10\leq N_e\leq 40$ and $\tilde{f}=3$ for the
        curvaton case in this figure.  (Notice that the data points
        for the standard case without the curvaton are unchanged.)}
        \label{fig:flow_ns_r_Cinf}
    \end{center}
    \begin{center}
        \centerline{\epsfysize=0.45\textwidth
        \epsfbox{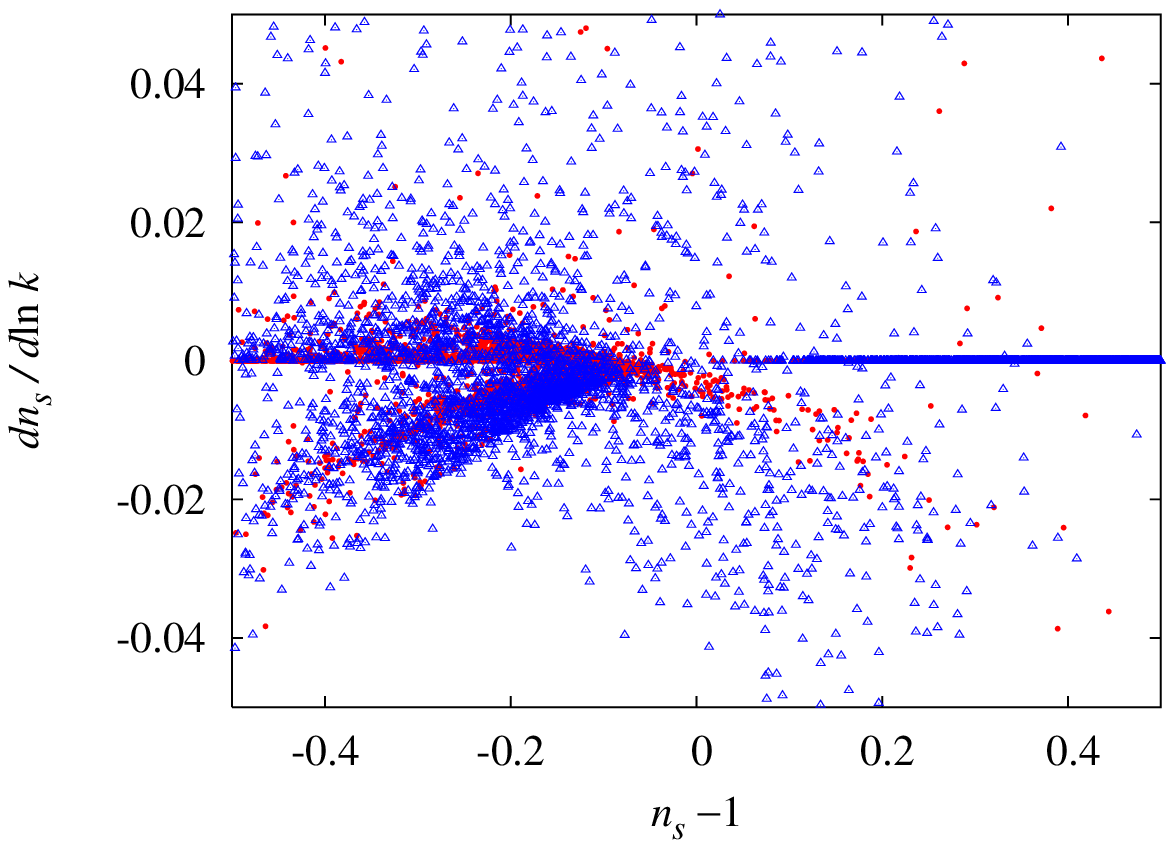}}
        \caption{The same as Fig.~\ref{fig:flow_ns_run} except 
        that we take $10\leq N_e\leq 40$ for the curvaton case in this
        figure.  }
        \label{fig:flow_ns_run_Cinf}
    \end{center}
\end{figure}

It is notable that the clustering structure shows up because of
relatively large value of $N_e$.  If a second inflation occurs with
the energy density of the curvaton, however, this result may be
changed.  With the second inflation, significant expansion after the
first inflation is possible and hence $N_e$ may drastically decrease.
When the initial amplitude of the curvaton becomes as large as $M_{\rm
pl}$, this can be the case.  Then, $N_e$ much smaller than $40-70$ may
be realized.  For some inflation models, reduced value of $N_e$ is too
small to make the point $(n_s-1,r)$ into the clustering region and,
consequently, the resulting distribution on the $n_s$ vs.\ $r$ plane
becomes more scattered than the previous case.  
 This fact has
some importance for models with $-4\epsilon_H+2\eta_H>1$ (which means
that, for {\it some} scale, the primordial spectrum is blue-tilted)
because the fixed point locates in the region where $n_s-1<0$.  For
models where the initial values of the slow-roll parameters satisfy the
relation $-4\epsilon_H+2\eta_H>1$, the value of $-4\epsilon_H+2\eta_H$
can be reduced to make $n_s -1\simeq 1$ in the course of approaching
to the fixed point.  As one can see, in the case with the curvaton,
larger number of points fall into the region consistent with the WMAP
data (i.e., $n_s\simeq 1$ with $r\lesssim 1$) in Fig.\ 
\ref{fig:flow_ns_r_Cinf} than in Fig.\ \ref{fig:flow_ns_r}.

To be more quantitative, we perform the numerical analysis with
smaller values of $N_e$.  The second inflation driven by the curvaton
may decrease the number of $e$-folding $\Delta N_e=20-30$.  Thus we
take the range of the initial $e$-folding number as $10\leq N_e\leq
40$ assuming the second inflation due to the curvaton.  (In fact, the
present horizon scale has to exit the horizon during the first
inflation, which gives an upper bound on the initial amplitude of the
curvaton.  Consequently, in the case with the second inflation, the
$\tilde{f}$ parameter cannot be so large; see Eq.\ (\ref{funf}).)  The
results with the mild value $\tilde{f}=3$ are shown in
Figs.~\ref{fig:flow_ns_r_Cinf} and \ref{fig:flow_ns_run_Cinf} in the
$n_s$ vs.\ $r$ and $n_s$ vs.\ $d\ln n_s/d\ln k$ planes, respectively.
As one can see, compared to the case with $40\leq N_e\leq 70$, the
distribution is more scattered for $10\leq N_e\leq 40$.  In
particular, sizable number of the inflation models are moved into the
region with $n_s\simeq 1$, the allowed region from the WMAP data,
which is not the case for the case without the second inflation.  In
fact, in the case with the second inflation, the change of the
distribution is mostly from the reduction of the $e$-folding number in
the first inflation.  Thus, the effects of the curvaton fluctuation
(i.e., the values of $\tilde{f}$) on the distribution of models in the
$n_s$ vs.\ $r$ plane is not so significant compared with the case with
small curvaton initial amplitude.

\section{Summary}

In this paper, we discussed the effects of the curvaton on inflation
models from some general points of view.  For this purpose, first we
classify inflation models in the $n_s$ vs.\ $r$ plane into three
categories: the ``small-field'' models, ``large-field'' models and
``hybrid-type'' models.

For the case that the initial amplitude of the curvaton is small
($\phi_{\rm init} < M_{\rm pl}$) and that the change of $N_e$ being
negligible, we have shown how the scalar spectral index and the
tensor-to-scalar ratio are modified for each models.  In the
``small-field'' models, both the spectral index and the
tensor-to-scalar ratio are not affected by the curvaton much (although
slight increase in $n_s$ can be seen).  In the ``large-field'' models,
the spectral index $n_s$ always increases and the tensor-to-scalar
ratio is largely suppressed.  In the ``hybrid-type'' models, the
spectral index always decreases which is the opposite effect compared
to the cases with the small-field and large-field models.

We also investigated the effects of the curvaton on inflation models
generated by the inflationary flow equation. Since we gave a general
argument how the observable quantities are affected by the curvaton,
we can easily understand the modification of the structure of
distribution of inflation models generated by the inflationary flow
equation.

{\it Acknowledgments:} T.T. would like to thank the Japan Society for
Promotion of Science for financial support.  The work of T.M. is
supported by the Grants-in Aid of the Ministry of Education, Science,
Sports, and Culture of Japan No.\ 15540247.

\end{document}